\documentclass{elsart}
\usepackage{amsmath,amsfonts,amssymb,graphicx}
\usepackage{mathrsfs,bbm}				%FOR MATHEMATICAL FONTS

\usepackage[sort&compress]{natbib}			%FOR BIBLIOGRAPHY
\bibpunct{[}{]}{,}{n}{}{;}				%FOR BIBSTYLE

	\renewcommand{\Re}{\ensuremath{\mathbb{R}}}

	\DeclareMathAlphabet{\mathpzc}{OT1}{pzc}{m}{it}
\begin{document}
\begin{frontmatter}
\title{Fractal Dimension of the Cantor Moir\'{e} Structures} 
\author{Luciano Zunino\corauthref{cor}}
\corauth[cor]{Corresponding author.}
\ead{lucianoz@ciop.unlp.edu.ar}
and
\author{Mario Garavaglia}
\ead{garavagliam@ciop.unlp.edu.ar}

\address{Centro de Investigaciones \'Opticas (CIOp), CC. 124 Correo Central,1900 La Plata, Argentina.}

\address{Departamento de F\'{\i}sica,
                 Facultad de Ciencias Exactas,
                 Universidad Nacional de La Plata (UNLP),
                 1900 La Plata, Argentina.}

%%%%%%%%%%%%%%%%%%%%%%%%%%%%%%%%%%%%%%%%%%%%%%%%%%%%%%%%%%%%%%%%%%%%%%%%%%%%%%%%%%%%%%%%%%%%%%%%%%%
%%%%%%%%%%%%%%%%%%%%%%%%%%%%%%%%%%%%%%%%%%%%%%%%%%%%%%%%%%%%%%%%%%%%%%%%%%%%%%%%%%%%%%%%%%%%%%%%%%%
%	ABSTRACT
%%%%%%%%%%%%%%%%%%%%%%%%%%%%%%%%%%%%%%%%%%%%%%%%%%%%%%%%%%%%%%%%%%%%%%%%%%%%%%%%%%%%%%%%%%%%%%%%%%%

\begin{abstract} 
In a recently published paper (J. of Modern Optics 50 (9) (2003) 1477-1486) a qualitative analysis 
of the moir\'{e} effect observed by superposing two grids containing Cantor 
fractal structures was presented. It was shown that the moir\'{e} effect is sensible to 
variations in the order of growth, dimension and lacunarity of the Cantor 
fractal. It was also verified that self-similarity of the original fractal 
is inherited by the moir\'{e} pattern. 

In this work it is shown that these Cantor moir\'{e} structures are also 
fractals and the fractal dimension associated with them is theoretically determined and experimentally measured attending the size of rhombuses in the different orders of growth. 
\end{abstract}

\end{frontmatter}

\section{INTRODUCTION}
\label{sec:intro}
\subsection{Fractals}
\label{sec:fractals}
In recent years the study of fractals has attracted a lot of attention 
because many physical phenomena and natural structures can be analysed and 
described by using a fractal approach \cite{book:mandelbrot82}. B. B. Mandelbrot introduced the 
term fractal, which comes from the Latin \textit{fractus} meaning broken, 
to describe irregular structures which are impossible to study using traditional 
Euclidean geometry. As he suggested, the principal characteristic of 
these new structures is their self-similarity: they are invariant under 
change of scale and displacement. A fractal is a set whose parts resemble 
the whole. This self-similarity can be mathematical or statistical. 
In the first case exact copies of the whole are obtained when the structure 
is viewed under magnification. These fractals are constructed through an 
iterative process of replacements of an object by $n$ copies of itself, each 
one of which is scaled by an $r < 1$ factor. Given a self-similar structure, 
there is a relation between the scaling factor $r$ and the number $n$ of 
pieces in which it is divided, according to the formula:
\begin{equation}
n = \frac{1}{r^{d_s }},
\end{equation}
where $d_s $ is the called similarity dimension. Fractal dimensions provide 
a description of how much space the set fills.

\begin{figure} 
\begin{center}
\includegraphics[width=0.8\textwidth]{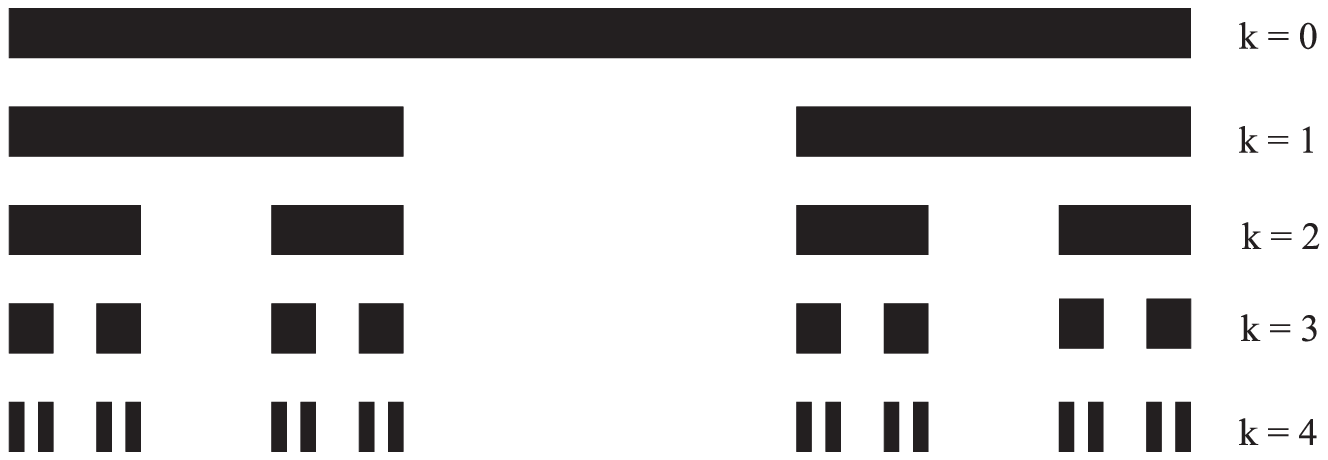}
\caption{Grids constructed following the first five orders of growth for the 
triadic Cantor fractal ($n = 2$ and $r = 1/3$).\label{figure1}}
\end{center}
\end{figure}

The classic triadic Cantor fractal is an example of these structures. In its 
build-up process a sequence of closed intervals is generated, one after the 
zero step, two after the first step, four after the second, eight after the 
third, and so on. In general there will be $2^k$ intervals of longitude $1 
\mathord{\left/ {\vphantom {1 {3^k}}} \right. \kern-\nulldelimiterspace} 
{3^k}$ after the $k^{th}$ step. Each iteration represents an order of 
growth. The Cantor fractal is defined as the array of points that remain 
after an infinite process of removals. In Fig. \ref{figure1} are shown grids 
constructed following the first five orders of the triadic Cantor fractal 
($n = 2$ and $r = 1 \mathord{\left/ {\vphantom {1 3}} \right. 
\kern-\nulldelimiterspace} 3)$. It has been drawn with a finite height to 
aid the viewing. 

\begin{center}
\textbf{[Insert figure 1 about here]}
\end{center}

A mathematical fractal is obtained by considering the structure that results 
when the order of growth $k \to \infty $. Practical fractals are 
self-similar over a limited range of magnification and they are more 
appropriately referred as pre-fractals. However, in this paper, the term 
fractal will be applied to structures with finite $k$.

Fractals, as it was mentioned, can also be obtained using a statistical 
process. The similarity dimension is not meaningful for this class of 
fractals. However, there are other definitions of dimension that are defined 
for any set. The Hausdorff dimension, which is based on measures, is the 
oldest and probably the most important. Box-counting or box dimension (also 
known as Kolmogorov entropy, entropy dimension, capacity dimension and 
information dimension) is another dimension very popular in 
physical applications.

In this work the focus is on strictly self-similar fractals. The Hausdorff 
dimension equals the similarity dimension for self-similar fractals \cite{book:feder88}.

\subsection{Moir\'{e} effect}
\label{sec:moire effect}
In order to characterize fractal structures, optical diffraction and scattering by 
fractal openings is being increasingly studied \cite{paper:allain86,paper:allain87,paper:zuninob03}. A complete list of references can be found in Ref.~\cite{paper:zuninob03}. It allows the properties and parameters that characterize these objects to be determined. 

\begin{figure} 
\begin{center}
\includegraphics[width=0.8\textwidth]{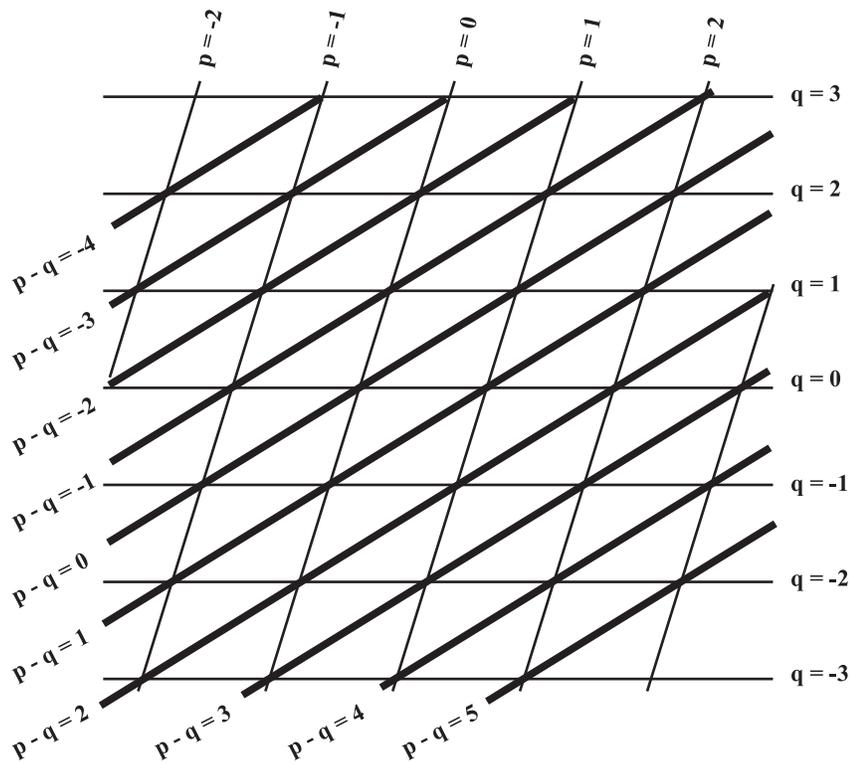}
\caption{Moir\'{e} pattern generated by superposition of two grids; $p$ and $q$ are the indices of both families of lines. Moir\'{e} fringes are represented by $p-q=0, \pm1, \pm2,...$ .\label{figure2}}
\end{center}
\end{figure}

More recently another powerful optical tool, the moir\'{e} effect, was 
applied to this particular geometry \cite{paper:zunino03,paper:calva02}. Moir\'{e} 
patterns are observed when two similar screens or sets of rulings are nearly superposed. 
They may be described as the locus of points 
of intersection of the two overlapping grids, as it is shown in Fig. \ref{figure2}. Assume that each of the two original grids can be regarded as an indexed family of lines. Then, the 
resulting moir\'{e} patterns are most pronounced when the indices of the 
intersections satisfy certain simple relations. This is known as the 
indicial representation method for the determination of the moir\'{e} 
patterns \cite{paper:oster64}, which could be called algebraic method. 

\begin{center}
\textbf{[Insert figure 2 about here]}
\end{center}

There are other methods to determine characteristics of moir\'{e} patterns using different approaches: geometric \cite{report:tolenaar45}, analytic \cite{paper:ronchi25}, vector \cite{paper:sciammarella82}, tensor \cite{paper:tatasciore95}, autocorrelation \cite{paper:alqazzaz75}, Fourier transformation \cite{paper:amidror94}, and by using a description of the superposed grids according to Walsh functions \cite{paper:colautti97}. These approaches could be a sequence initiated as early as 1874 by Lord Rayleigh --who was the first to explain how moir\'{e} patterns are observed from the superposition of two families 
of equispaced parallel straight lines \cite{paper:strutt74}--, and continues to date.

A Ronchi grid can be described as $G\left( {x,y,d,\phi } \right)$ by the 
expression:
\begin{equation}
G\left( {x,y,d,\phi } \right) = \sum\limits_{n = 1}^N {rect\left\{ 
{\frac{\left[ {x - nd - \phi (x,y)} \right]}{\raise0.7ex\hbox{$d$} 
\!\mathord{\left/ {\vphantom {d 
2}}\right.\kern-\nulldelimiterspace}\!\lower0.7ex\hbox{$2$}}} \right\}} ,
\end{equation}
where $x$ is the variable over which the rectangular function describes the Ronchi grid distribution, $y$ the variable that permits describing the grid $G\left( {x,y,d,\phi } \right)$ in the plane $\left( {x,y} \right)$, 
$d$ the grid period and $\phi $ the phase of the grid considered as its 
first line position respect to the left border of the reference frame. 
Formally, the bidimensional Ronchi grid can be represented as the Cartesian 
product $G = ExR$, being $E$ a straight line segment parallel to the y-axis 
of coordinates and $R$ the rectangular function in the x-axis of coordinates. Different operators can be applied to the function $G\left( 
{x,y,d,\phi } \right)$, such as translation $T\left( {x,y} \right)$ over the 
plane $\left( {x,y} \right)$, scaling $S\left( {x,y} \right)$ that produces 
a variation of the original period $d$, and rotation $A\left( \alpha 
\right)$ of the grid around its $z$ axis in a certain angle $\alpha $.

Finally, the superposition of the original grid $G\left( {x,y,d,\phi } 
\right)$ with the modified grid $G'\left( {x,y,d,\phi } \right)$, generated 
by the application of the operators $T\left( {x,y} \right)$, $S\left( {x,y} 
\right)$, and $A\left( \alpha \right)$, or any of them, over $G\left( 
{x,y,d,\phi } \right)$, allows determining, for example, the transmittance 
$t\left( {x,y} \right)$ of the moir\'{e} pattern by means of the correlation 
$G * G'$, according to:
\begin{equation}
t\left( {x,y} \right) = \int\limits_{ - \infty }^\infty {\int\limits_{ - 
\infty }^\infty {G\left( {x,y,d,\phi } \right)} } G'\left( {x,y,d,\phi } 
\right)dxdy.
\end{equation}

\begin{figure} 
\begin{center}
\includegraphics[width=0.55\textwidth]{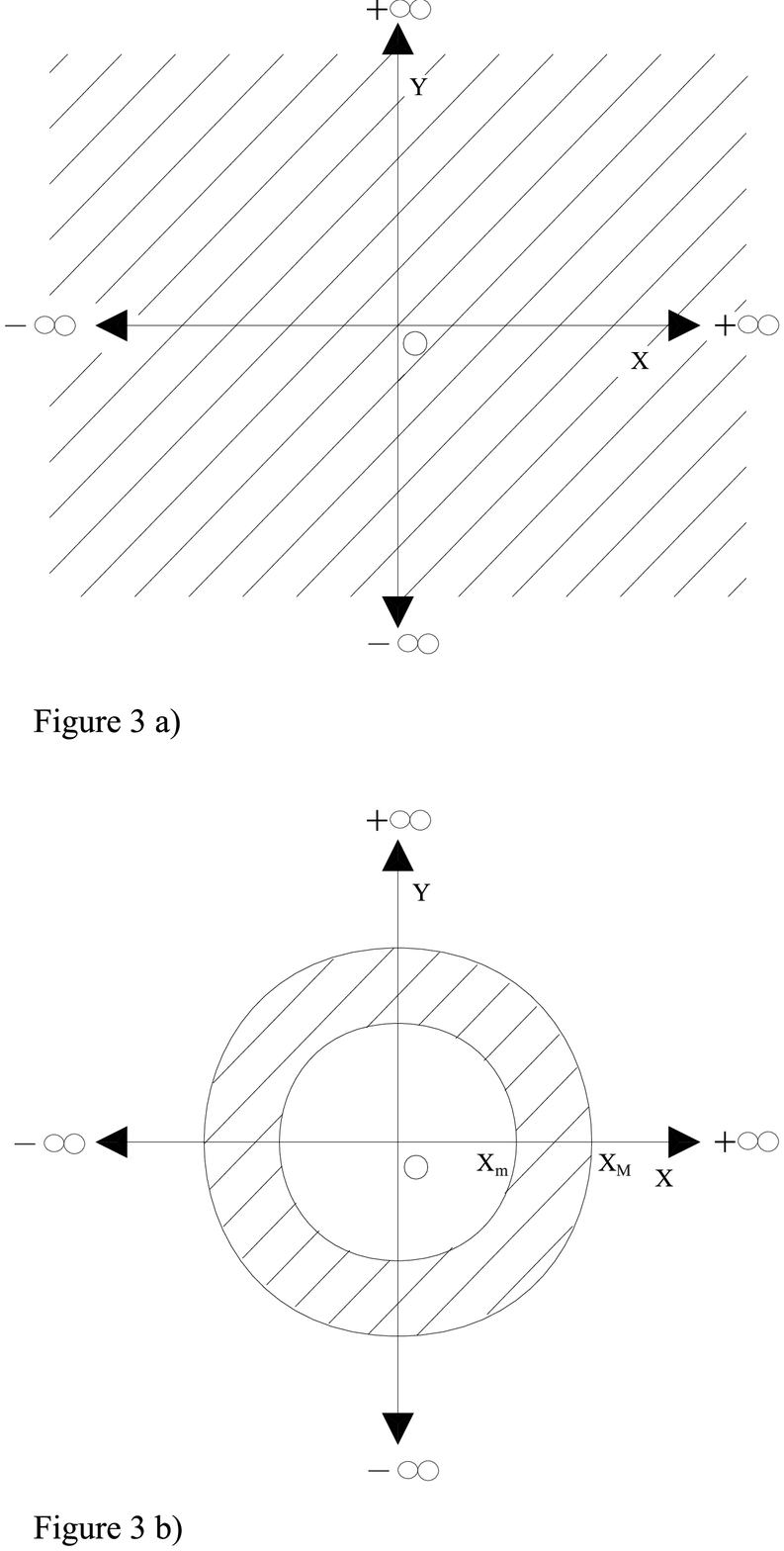}
\caption{a) The bidimensional correlation function is defined in 
such a way that must be integrated over the entire plane, from $ - \infty $ 
to $ + \infty $. b) The bidimensional correlation function for detectors is 
limited by the macroscopic dimensions of the devices (X$_{M})$ and the 
microscopic dimensions of its sensible components (X$_{m})$.\label{figure3}}
\end{center}
\end{figure}

When referring to the appearance of the moir\'{e} phenomenon, the verbal 
form ``visual'' is often used to describe the observation of the emerging 
geometrical figures called moir\'{e} patterns. It seems to be a little 
reductionistic approach to describe it, because not only eyes observe 
moir\'{e} patterns. In fact, any type of image capture system can be 
successfully utilized to capture moir\'{e} patterns and to display them to 
be visually observed. Actually, moir\'{e} patterns can be photographed, 
photocopied, taken by TV camera, PC designed and scanned, and, after some 
appropriate processes is applied to the captured information, they can be 
visually observed. Different types of display support are employed, as paper 
reproduced pictures, projected slides or transparencies, photocopies, paper 
printed images, TV and monitor screen images, etc. However, it is convenient 
to mention that all detectors and displays listed above are discrete in 
nature and finite in size. They have their proper structure. Then, the 
bidimensional correlation function for detectors is limited by the 
macroscopic dimensions of the devices (X$_{M})$ and the microscopic 
dimensions of its sensible components (X$_{m})$ as shown in Fig. \ref{figure3}.

\begin{center}
\textbf{[Insert figure 3 about here]}
\end{center}

Then the best matching between moir\'{e} patterns and the structure of 
detectors must be accomplished to assure the clearest observation of them 
and avoid the appearance of the noisy aliasing effect \cite{paper:bell84}. Also, the best 
observation of moir\'{e} fringes is strongly limited by the poor capacity of 
the eye to distinguish very low and very high spatial frequency components 
in an image. Finally, the observation of moir\'{e} patterns is related with 
local and global correlations of signals in the visual system \cite{book:marr82}.

\begin{figure} 
\begin{center}
\includegraphics[width=0.35\textwidth]{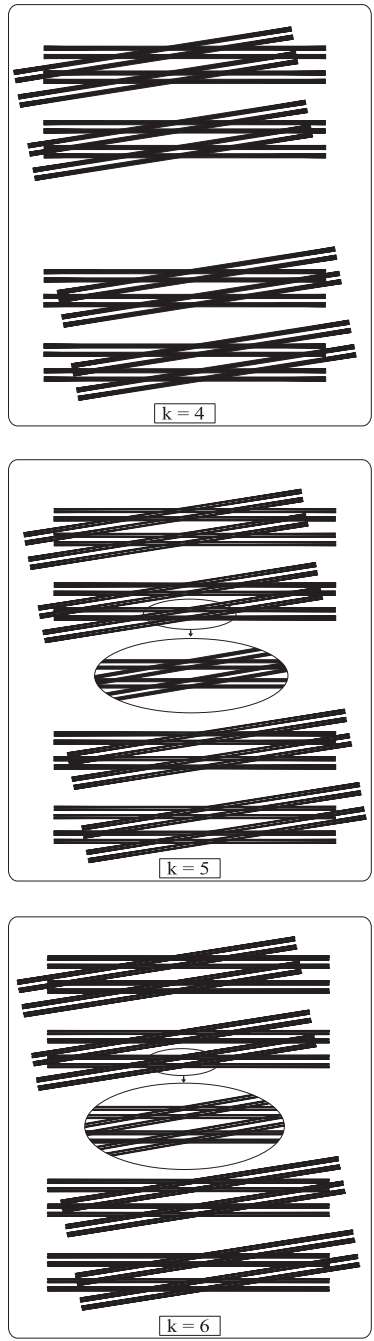}
\caption{Moir\'{e} patterns superposing triadic Cantor fractals 
($n = 2$ and $r = 1/3$) for the orders of growth $k = 4$ (upper figure), 
$k = 5$ (central figure), and $k = 6$ (lower figure).\label{figure4}}
\end{center}
\end{figure}

Moir\'{e} patterns are obtained from different types of $2D$ grids: 
equispaced parallel lines---as in Fig. \ref{figure2}---, parallel lines of variable spacing, radial lines, circles whose difference between consecutives radii is constant, zone 
plates, parabolas, spirals, etc. It has been shown that the moir\'{e} effect 
is also present after superposing a Cantor bidimensional structure over its 
own replica rotated in a small angle. It was also verified that 
self-similarity of the original fractal is inherited by the moir\'{e} 
pattern \cite{paper:zunino03}. In order to illustrate this property, Fig. \ref{figure4} shows the 
moir\'{e} fringes obtained by superposing the triadic Cantor fractals of 
orders $k = 4$, $k = 5$ and $k = 6$. In all the moir\'{e} figures the 
fractal structures are superposed over their replicas, which have been 
rotated an angle of 10$^{o}$. It is observed that the moir\'{e} patterns 
contain tinier and more complex structures as the order of growth increases. 
The central parts of the moir\'{e} patterns for orders of growth $k = 5$ and 
$k = 6$ were magnified to observe the structure in detail. The features in 
the magnified region directly correspond to the characteristic moir\'{e} 
features of the precedent order. So the moir\'{e} patterns inherit the 
self-similarity of the original Cantor structures.

\begin{center}
\textbf{[Insert figure 4 about here]}
\end{center}

Now, in this work it will be analysed if these Cantor moir\'{e} structures 
are fractals. To our knowledge this problem has not been introduced before. 

The use of the term triadic Cantor fractal is, strictly, an abuse of 
language, since in fact what is superposed is the structure which results of 
extending the triadic Cantor perpendicularly within the $2D$ plane. However, 
it is a $1D$ geometry because it only has full freedom in one dimension, 
while its other dimension is completely determined. Then, it is possible to 
consider it as the Cartesian product $C = ExF$:
\begin{equation}
C = ExF = \left\{ {\left( {x,y} \right):x \in E,y \in F} \right\},
\end{equation}
where $E$ is a straight line segment and $F$ is the triadic Cantor set. It 
was shown \cite{book:falconer90} that this structure has Hausdorff dimension:
\begin{equation}
\dim _H C = \dim _H \left( {ExF} \right) = \dim _H E + \dim _H F = 1 + 
\frac{\ln 2}{\ln 3}.
\end{equation}

\section{Analysis}
\label{sec:analysis}
The superposition of bidimensional structures to generate the moir\'{e} 
effect is expressed by the logical operation of intersection $ \cap $, which 
can be quantified by means of the analytical operation of correlation of the 
mathematical functions that describe the structures.

\begin{figure} 
\begin{center}
\includegraphics[width=1\textwidth]{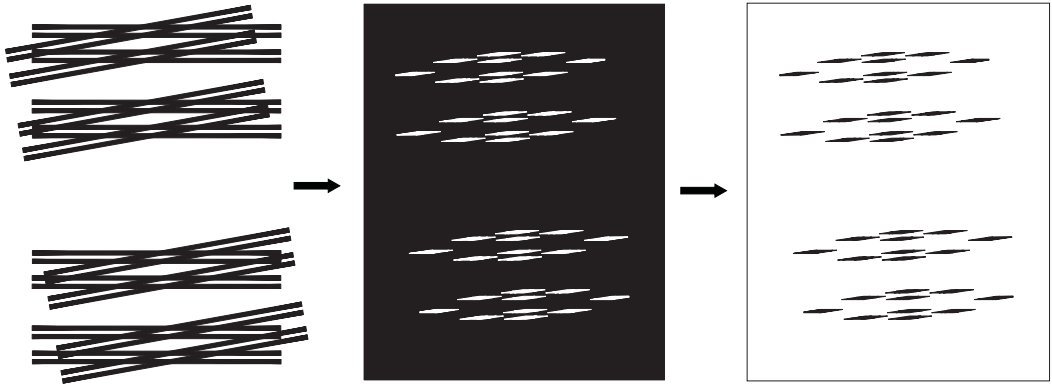}
\caption{Operations introduced in order to visualize the 
intersection regions (triadic Cantor fractal, order of growth $k = 4)$.\label{figure5}}
\end{center}
\end{figure}

In order to visualize the intersection between the two Cantor moir\'{e} 
structures the following elementary algebra concept is applied:
\begin{equation}
A \cap B = \left( {A^c \cup B^c} \right)^c.
\end{equation}
The moir\'{e} structure corresponds to the complement region obtained from 
the union of the complement of the original structures. In Fig. \ref{figure5} it is 
illustrated this operation for the triadic Cantor fractal with order of 
growth $k = 4$.

\begin{center}
\textbf{[Insert figure 5 about here]}
\end{center}

\begin{figure} 
\begin{center}
\includegraphics[width=0.9\textwidth]{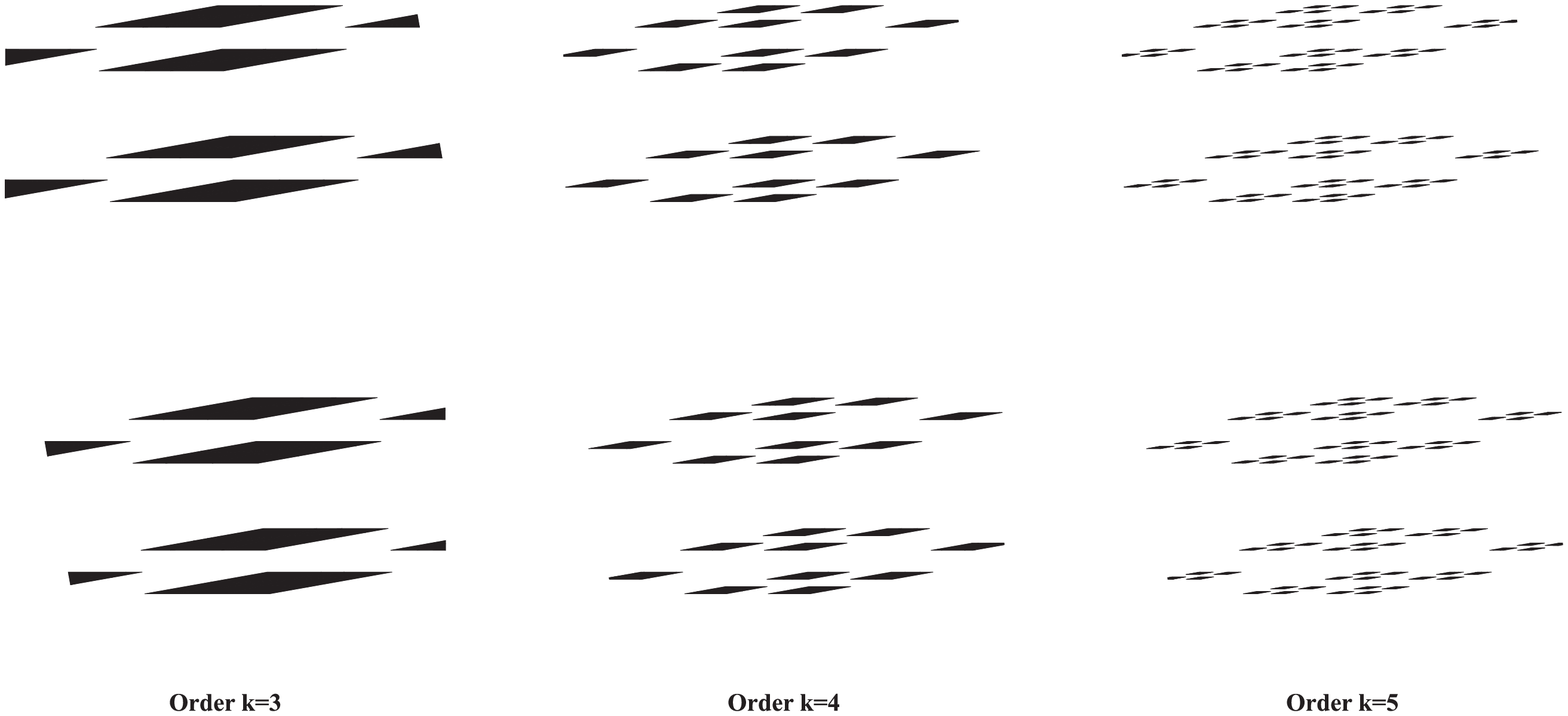}
\caption{Intersection regions of the triadic Cantor fractals for 
the orders of growth $k = 3$ (left figure), $k = 4$ (central figure), and $k = 
5$ (right figure).\label{figure6}}
\end{center}
\end{figure}

Now, it is easy to follow the behaviour of the moir\'{e} when the order of 
growth is increased. Figure \ref{figure6} shows the intersection regions for the orders 
of growth $k = 3$, $k = 4$ and $k = 5$.

\begin{center}
\textbf{[Insert figure 6 about here]}
\end{center}

As the order of growth is increased the original rhombuses that result from 
the intersection are divided in new four small rhombuses with length size 
scaled by a factor $r = 1 \mathord{\left/ {\vphantom {1 3}} \right. 
\kern-\nulldelimiterspace} 3$. So, it is possible to conclude that the 
similarity dimension of the moir\'{e} structures is:
\begin{equation}
d_s = \frac{\ln 4}{\ln 3} = 2\frac{\ln 2}{\ln 3}.
\end{equation}
A justification of this result can be achieved by analysing the intersection 
formula for fractals. The following two theorems are used \cite{book:falconerb90}:

\textit{Theorem A}. If $A$, $B$ are Borel subsets of $\Re^n$ and $\sigma $ ranges over a group $G$ of transformations, such as the group of translations, congruences or 
similarities then:
\[
\dim _H \left( {A \cap \sigma \left( B \right)} \right) \le \max \left\{ 
{0,\dim _H \left( {AxB} \right) - n} \right\},
\]
for almost all $x \in \Re ^n$.

\textit{Theorem B}. Let $A,B \subset \Re ^n$ be Borel sets, and let $G$ be a group of 
transformations on $\Re ^n$. Then:
\[
\dim _H \left( {A \cap \sigma \left( B \right)} \right) \ge \dim _H A + \dim 
_H B - n,
\]
for a set of motions $\sigma \in G$ of positive measure in the following 
cases:

(a) $G$ is the group of similarities and $A$ and $B$ are arbitrary sets;

(b) $G$ is the group of rigid motions, $A$ is arbitrary and $B$ is a 
rectifiable curve, surface, or manifold;

(c) $G$ is the group of rigid motions and $A$ and $B$ are arbitrary, with 
either $\dim _H A > \frac{1}{2}\left( {n + 1} \right)$ or $\dim _H B > 
\frac{1}{2}\left( {n + 1} \right)$.

Remember that a rigid motion or direct congruence may be achieved by a 
combination of a rotation and a translation. It does not involve reflection.

The Cantor moir\'{e} structures have $A = B = C = ExF$, $n = 2$ and $\sigma 
$ a rotation. Then:
\begin{equation}
\dim _H \left( {C \cap \sigma \left( C \right)} \right) \le \max \left\{ 
{0,\dim _H \left( {CxC} \right) - 2} \right\},
\end{equation}
and
\begin{equation}
\dim _H \left( {C \cap \sigma \left( C \right)} \right) \ge 2\dim _H C - 2 = 
2\left( {1 + \frac{\ln 2}{\ln 3}} \right) - 2 = 2\frac{\ln 2}{\ln 3}.
\end{equation}

But $\dim _H \left( {CxC} \right) = 2\dim _H C = 2\left[ {1 + \left( {{\ln 
2} \mathord{\left/ {\vphantom {{\ln 2} {\ln 3}}} \right. 
\kern-\nulldelimiterspace} {\ln 3}} \right)} \right]$ \cite{book:falconerc90}, so the following 
upper bound is obtained:
\begin{equation}
\dim _H \left( {C \cap \sigma \left( C \right)} \right) \le \max \left\{ 
{0,\dim _H \left( {CxC} \right) - 2} \right\} = \max \left\{ {0,2\frac{\ln 
2}{\ln 3}} \right\} = 2\frac{\ln 2}{\ln 3}\hspace{1em}.
\end{equation}
Then, according to the results in equation (9) and (10) the Hausdorff 
dimension of the Cantor moir\'{e} structures straightforward equals to 
$2\hspace{2pt}{\ln 2} \mathord{\left/ {\vphantom {{\ln 2} {\ln 3}}} \right. 
\kern-\nulldelimiterspace} {\ln 3}$.

\begin{figure}
\begin{center}
\includegraphics[width=0.8\textwidth]{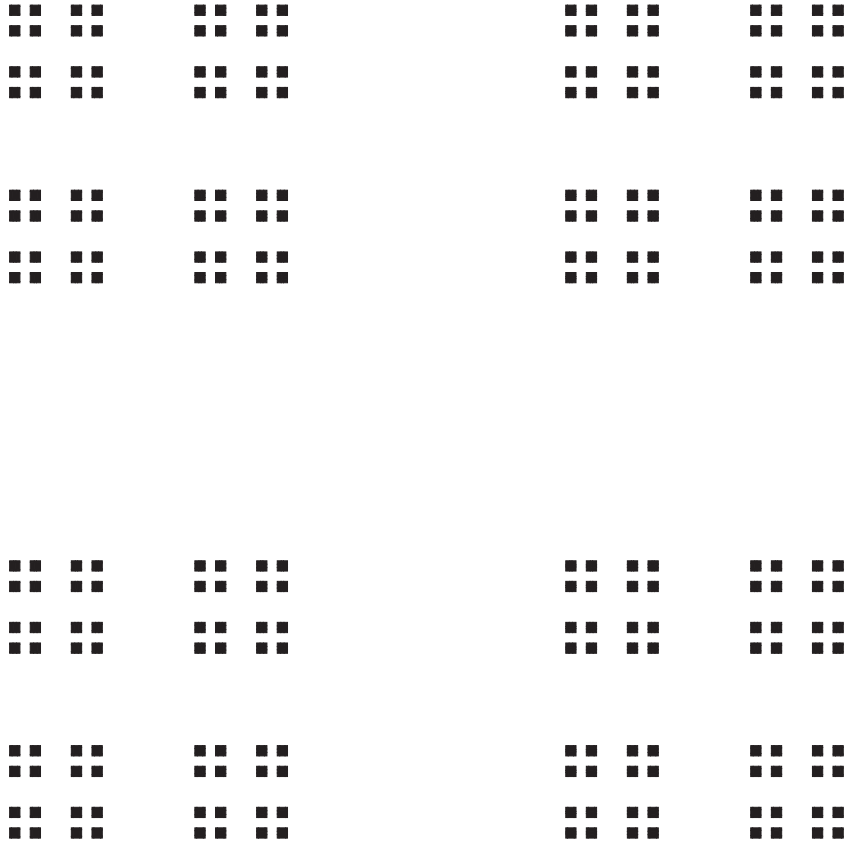}
\caption{Cartesian product of the triadic Cantor fractal with 
itself for the order of growth $k = 4$.\label{figure7}}
\end{center}
\end{figure}

It is possible another explanation of this result by analysing the Cartesian 
product $F\times F$, where $F$ is the triadic Cantor fractal. Figure \ref{figure7} shows 
this product. It was shown \cite{book:falconerd90} that this Cartesian Cantor product has a 
Hausdorff dimension exactly $2\hspace{2pt}{\ln 2} \mathord{\left/ {\vphantom {{\ln 2} 
{\ln 3}}} \right. \kern-\nulldelimiterspace} {\ln 3}$.

\begin{center}
\textbf{[Insert figure 7 about here]}
\end{center}

The Cantor moir\'{e} structure is obtained under a bi-Lipschitz 
transformation $f$ of this Cartesian product, i.e.:
\begin{equation}
f:X \to Y,
\quad
c_1 \left| {x - y} \right| \le \left| {f\left( x \right) - f\left( y 
\right)} \right| \le c_2 \left| {x - y} \right|,
\quad
\left( {x,y \in X} \right),
\end{equation}
for $0 < c_1 \le c_2 < \infty $. Furthermore, the Hausdorff dimension is 
invariant under bi-Lipschitz transformation \cite{book:falconere90}. Then, it is confirmed that 
the Hausdorff dimension of the new structure is $2\hspace{2pt}{\ln 2} \mathord{\left/ 
{\vphantom {{\ln 2} {\ln 3}}} \right. \kern-\nulldelimiterspace} {\ln 3}$.

\begin{figure}
\begin{center}
\includegraphics[width=0.9\textwidth]{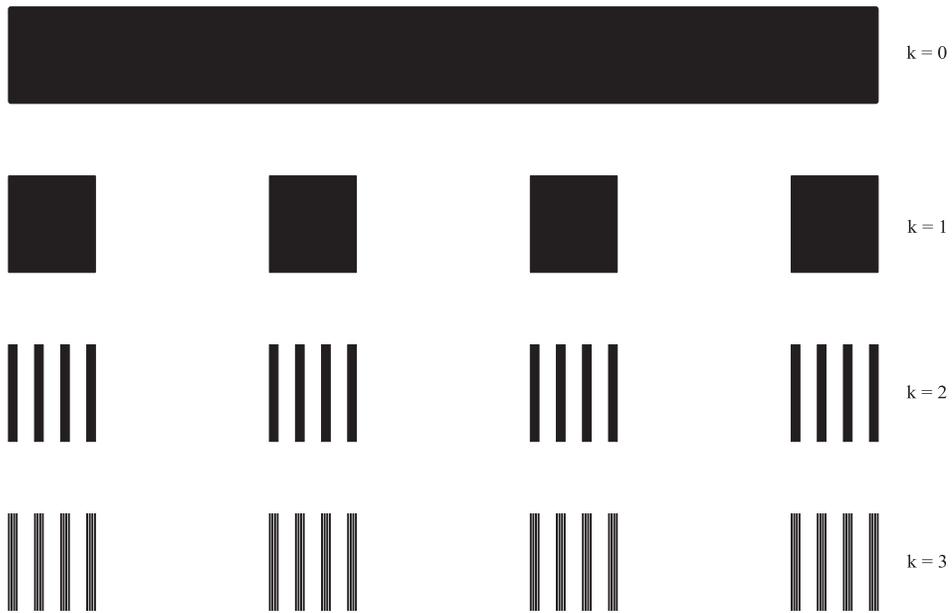}
\caption{Grids constructed following the first orders of growth of 
the septic Cantor fractal ($n = 4$ and $r = 1/7$).\label{figure8}}
\end{center}
\end{figure}

It is easy to extend these results to other Cantor sets. Figure \ref{figure8} shows the 
septic Cantor bars ($n = 4$ and $r = 1 \mathord{\left/ {\vphantom {1 7}} 
\right. \kern-\nulldelimiterspace} 7$, being $d = {\ln 4} \mathord{\left/ 
{\vphantom {{\ln 4} {\ln 7}}} \right. \kern-\nulldelimiterspace} {\ln 7})$ 
in their first four orders of growth.

\begin{center}
\textbf{[Insert figure 8 about here]}
\end{center}

\begin{figure}
\begin{center}
\includegraphics[width=1\textwidth]{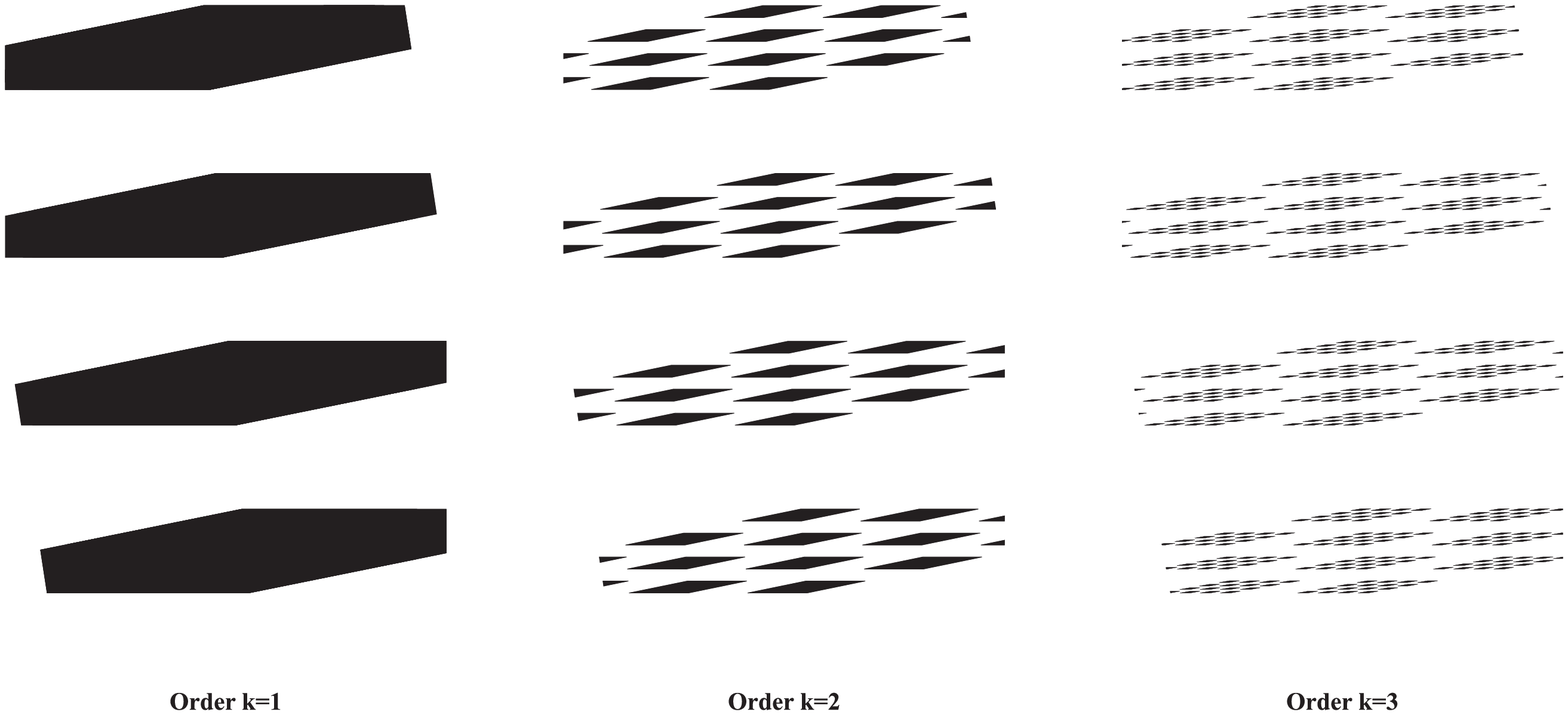}
\caption{Intersection regions of the septic Cantor fractals for 
the orders of growth $k = 1$ (left figure), $k = 2$ (central figure), and $k = 
3$ (right figure).\label{figure9}}
\end{center}
\end{figure}

The intersection regions associated to this fractal for the orders of growth 
$k = 1$, $k = 2$ and $k = 3$ are shown in Fig. \ref{figure9}. The original rhombuses 
are divided in new sixteen small rhombuses with length size scaled by a 
factor $r = 1 \mathord{\left/ {\vphantom {1 7}} \right. 
\kern-\nulldelimiterspace} 7$. Then, 
\begin{equation}
d_s = \frac{\ln 16}{\ln 7} = 2\frac{\ln 4}{\ln 7},
\end{equation}
as it was expected.

\begin{center}
\textbf{[Insert figure 9 about here]}
\end{center}

\section{Conclusions}
\label{sec:conclusions}
It can be concluded that the Cantor moir\'{e} structures are self-similar 
fractals with twice the similarity and Hausdorff dimension of the original 
Cantor structures.

The extension of these results to other fractal constructions is of great 
importance because new fractal structures can be obtained. The intention is 
to demonstrate that one way of constructing `new fractals from old' is by 
forming moir\'{e} with them\footnote{It is reproduced the following phrase 
``One way of constructing `new fractals from old' is by forming Cartesian 
products'' \cite{book:falconerf90} but modified for the new situation.}.

\ack   
 
Luciano Zunino thanks for the doctoral research fellowship from Consejo Nacional de Investigaciones Cient\'{\i}ficas y T\'ecnicas (CONICET), Argentina, during the research period.

\bibliography{references}
\bibliographystyle{elsart-num}

\newpage

\end{document}